\documentclass[showpacs,showkeys,aps,prd,floatfix,amsfonts,amsmath,amssymb]{revtex4}
\usepackage{amssymb}
\usepackage{amsmath}
\usepackage{graphicx}
\usepackage{bm}

\begin{document}

\title{Second order formalism for spin $\frac{1}{2}$ fermions and Compton scattering.}
\author{E. G. Delgado-Acosta, Mauro Napsuciale}
\affiliation{Departamento de F\'{\i}sica, Universidad de Guanajuato, Lomas del Bosque 103, Fraccionamiento Lomas del Campestre, C. P. 37150, Le\'{o}n, Guanajuato, M\'{e}xico.}
\author{Sim\'{o}n Rodr\'{\i}guez}
\affiliation{ Facultad de Ciencias F\'{\i}sico Matem\'{a}ticas, 
Universidad Aut\`{o}noma de Coahuila, Edificio "D", 
Unidad Camporredondo, CP 25280, Saltillo, Coahuila, M\'{e}xico.}
\begin{abstract}
We develop a second order formalism for spin 1/2 fermions based
on the projection over Poincar\'{e} invariant subspaces in the $(\frac{1}%
{2},0)\oplus(0,\frac{1}{2})$ representation of the homogeneous Lorentz group. Using 
$U(1)_{em}$ gauge principle we obtain second order description 
for the electromagnetic interactions of a spin 1/2
fermion with two free parameters, the gyromagnetic factor $g$ and a parameter $\xi$ 
related to odd-parity Lorentz structures.  We calculate Compton
scattering in this formalism. In the particular case
$g=2,~\xi=0$ and for states with well defined parity we recover Dirac results. 
In general, we find the correct classical 
limit and a finite value $r_{c}^{2}$ for the forward differential cross section, independent of the photon energy and of the value of the parameters $g$ and $\xi$. 
The differential cross section vanishes at high energies 
for all $g,~\xi$ except in the forward direction.  The total cross section at
high energies vanishes only for $g=2,~\xi=0$. We argue that this formalism 
is  more convenient than Dirac theory in the description of low energy 
electromagnetic properties of baryons and illustrate the point with the proton case. 
\end{abstract}
\maketitle

\section{ Introduction}

States describing a free particle transform in the irreducible representations
(irreps) of the Poincar\'{e}\ group. We will denote these states as
$\left\vert \Gamma\right\rangle $ where $\Gamma$ generically denotes the set
of good quantum numbers of this group among which we have the quantum numbers
associated to the Casimir operators, the squared four-momentum operator
$P^{2}$ and the squared Pauli-Lubanski operator $W^{2}$. These states can be
obtained from the vacuum using the creation and annihilation operators as
\begin{equation}
\left\vert \Gamma\right\rangle =a^{\dagger}(\Gamma)\left\vert 0\right\rangle .
\end{equation}
The transformation properties of the creation and annihilation operators under
Poincar\'{e}\ group are fixed by this relation. Poincar\'{e}\ invariance of
the scattering matrix and cluster decomposition implies the following general
form for the fields \cite{Weinberg:1995mt}
\begin{equation}
\psi_{l}(x)=\int d\Gamma\lbrack\kappa e^{-ip\cdot x}u_{l}(\Gamma
)a(\Gamma)+\lambda e^{ip\cdot x}v_{l}(\Gamma)a^{c\dagger}(\Gamma
)],\label{eq:fieldD}%
\end{equation}
and $\psi_{l}(x)$ must transform\textbf{ }in some representation of the
Homogeneous Lorentz Group\ (HLG) which in turn requires (because of the fixed
transformation properties of the creation operators) the coefficients
$u_{l}(\Gamma)$ and $v_{l}(\Gamma)$ also to transform in this representation
of the HLG\ . In this construction, extensively discussed in
\cite{Weinberg:1995mt} and which we will adopt here, the coefficients
$u_{l}(\Gamma)$ and $v_{l}(\Gamma)$ are just the Clebsch-Gordon coefficients
to select from the chosen HLG representation the desired
Poincar\'{e}\ irrep. The case of particles with well defined parity in the
$(\frac{1}{2},0)\oplus(0,\frac{1}{2})$ representation of the HLG can be found
in \cite{Weinberg:1995mt} and yields the conventional Dirac formalism where
the coefficients $u_{l}(\Gamma)$ and $v_{l}(\Gamma)$ are now interpreted as
states with well defined parity. Form this point of view, Dirac theory is
appropriate for the description of particles with well defined parity and it
can be shown that Dirac equation is just the equation for projection onto well
defined states in the $(\frac{1}{2},0)\oplus(0,\frac{1}{2})$ representation.
Similarly, the conventional Proca equation is just the projection onto a
subspace with well defined parity in the $(\frac{1}{2},\frac{1}{2})$
representation of the HLG. Unlike the $(\frac{1}{2},0)\oplus(0,\frac{1}{2})$
representation which contains only spin 1/2, in the 
$(\frac{1}{2},\frac{1}{2})$  case there are two
Poincar\'{e} sectors and it can be shown that parity projection and
Poincar\'{e} projection coincide.

In \cite{Napsuciale:2006wr} a formalism was developed for the description of spin 3/2 fields
using the projectors over well defined mass and spin in the spinor-vector
representation. These ideas were also applied to the vector representation in
\cite{Napsuciale:2007ry} . Gauging the formalism to account for electromagnetic
interactions, in both cases we obtain a theory containing a free parameter,
$g$, which can be identified with the gyromagnetic factor. For elementary
particles, this parameter is fixed using general arguments. In the spin 3/2
case it is shown in \cite{Napsuciale:2006wr} that causality of the interacting theory
requires $g_{\frac{3}{2}}=2$ at tree level. This value for the gyromagnetic
factor has been obtained for general fields using unitarity, gauge invariance
and rotational symmetry arguments \cite{Weinberglectures} (see also \cite{Ferrara:1992yc},\cite{Holstein:2006wi},\cite{Holstein:2006ry}), but to the best 
of our knowledge has never been related to causality of the interacting theory which
historically has been a severe problem for interacting high spin fields. The
Poincar\'e projector formalism was used to calculate Compton scattering
off spin 3/2 particles \cite{DelgadoAcosta:2009ic}. It was shown there that 
$g_{\frac{3}{2}}=2$ is necessary for the good behavior of differential cross section for
Compton scattering off spin 3/2 particles in the forward direction. Concerning
vector fields, the good behavior of the total cross section for Compton
scattering at high energies also requires $g_{1}=2$ \cite{Napsuciale:2007ry}. 
It can be shown that this value is required also by the good behavior of the differential 
cross section
in the forward direction. With this value, the Poincar\'{e} projector
formalism predicts the same electromagnetic couplings for the $W$ as those
obtained in the Standard Model.

As a side result, since the electromagnetic couplings of vector and
spinor-vector fields are described by a single parameter beyond the electric
charge, the Poincar\'{e} projector formalism predicts specific relations
between the multipole moments of spin 1 particles transforming in the vector
representation and spin 3/2 particles transforming in the spinor-vector
representation of the HLG. In the vector case, these relations are explicitly
given in \cite{Napsuciale:2007ry} and in particular are well satisfied by the 
$W$ boson whose magnetic dipole and electric quadrupole moments have been measured 
by the DELPHI Coll. \cite{Abreu:2001rpa}. A closer look at these relations will be
presented elsewhere.

In this work, we will be interested in developing a formalism for the
description of spin 1/2 particles using Poincar\'{e} projectors. These
projectors fix only the properties defining a particle, mass and spin, but do
not fix any other property such as parity thus we expect a proper description
of general spin 1/2 particles, not only Dirac particles (spin 1/2 particles
with well defined parity). Poincar\'{e} Projectors yield second order
equations thus we obtain a second order formalism for spin 1/2
fermions. Such possibility was addressed by Feynman and Gellmann long ago
aiming to describe chiral fermions required by the weak interactions
\cite{Feynman:1958ty}. Some authors studied a second order formalism for fermions 
along Feynman-Gellmann proposal \cite{Brown:1958zz} \cite{Tonin:1959} \cite{Barut:1962}
 \cite{Biedenharn:1972ih}\cite{CufaroPetroni:1985tu}. 
We will see below that the projector formalism differ from Feynman-Gellmann like formalisms.

The structure of this work is the following: in the next section we give all
the elements concerning the solving of the HLG algebra for the $(\frac{1}%
{2},0)\oplus(0,\frac{1}{2})$ representation, section III is devoted to the construction of
the formalism, in section IV we calculate Compton scattering and compare with
some experimental results for Compton scattering off protons at low energies. We summarize our results in section V.

\section{Solving the HLG algebra for $(\frac{1}{2},0)\oplus(0,\frac{1}{2})$.}

By ``solving the algebra" we mean the construction of all the algebraic
elements of a given representation, including states, from the HLG algebra,
without using any other argument such as equations of motions, lagrangian etc.
As well known, the generators of the HLG are the rotation and boost generators
$\{\mathbf{J},\mathbf{K}\}$ which satisfy the following algebra%
\begin{equation}
\lbrack J_{i},J_{j}]=i\epsilon_{ijk}J_{k},\qquad\lbrack J_{i},K_{j}%
]=i\epsilon_{ijk}K_{k},\qquad\lbrack K_{i},K_{j}]=-i\epsilon_{ijk}J_{k}.
\end{equation}
The part of the HLG connected to the identity is isomorphic to the
$SU(2)_{A}\otimes SU(2)_{B}$ group generated by the operators%
\begin{equation}
\mathbf{A}=\frac{1}{2}(\mathbf{J}-i\mathbf{K}),\qquad\mathbf{B}=\frac{1}%
{2}(\mathbf{J}+i\mathbf{K}), \label{AB}%
\end{equation}
hence the irreps of the HLG can be characterized by two independent $SU(2)$
quantum numbers $(a,b)$. A given irrep $(a,b)$ has dimension $(2a+1)(2b+1)$
and the states in this irrep are labeled by the corresponding quantum numbers
$|a,m_{a};b,m_{b}\rangle$ where $m_{a}$ and $m_{b}$ are the eigenvalues of
$A_{3}$ and $B_{3}$ respectively. The irreps with well defined value of
$\mathbf{J}^{2}$ are those with $a=0$ or $b=0$. In the case $b=0$ the
representations of the rotations and boost generators are related as
$\mathbf{J}=-i\mathbf{K}$ and since $\mathbf{A}=\mathbf{J}$ we denote these
irreps as $(j,0)$ and refer to them as ``right" representations of spin $j$ .
In the case $a=0$ we get $\mathbf{J}=i\mathbf{K}$, thus $\mathbf{B}%
=\mathbf{J}$ and we denote these irreps as $(0,j)$ and refer to them as
``left" representations of spin $j$ . Since we know how to construct a
representation for the $SU(2)$ rotation group, in both cases we have a
representation for the boost operator and it is possible to explicitly
construct the states in the basis $|j,m\rangle$ of well defined $\mathbf{J}%
^{2}$ and $J_{3}$ starting with the rest frame states. Explicitly, the boost
operators for these representations are%
\begin{align}
B_{R}(\mathbf{p)}  &  =\mathbf{\exp(-}i\mathbf{K}\cdot\mathbf{n}%
\varphi)=\mathbf{\exp(J}\cdot\mathbf{n}\varphi),\label{BRBL}\\
B_{L}(\mathbf{p)}  &  =\mathbf{\exp(-}i\mathbf{K}\cdot\mathbf{n}%
\varphi)=\mathbf{\exp(-J}\cdot\mathbf{n}\varphi).\nonumber
\end{align}
If we apply the boost operator on a rest frame state with momentum $p^{\mu
}=(m,\mathbf{0})$, then the rapidity $\varphi$ is related to the energy and
momentum of the particle as%

\begin{equation}
\cosh\varphi=\gamma=\frac{E}{m},\quad\sinh\varphi=\gamma\beta=\frac
{|\mathbf{p}|}{m}.
\label{rap}
\end{equation}
Using the explicit representation of the $\mathbf{J}$ operators it is
possible to construct a representation for the states in momentum space and in
the $\{|j,m\rangle\}$ basis.

Let us consider first the $(\frac{1}{2},0)$ case. In the conventional basis
$|\frac{1}{2},m\rangle$ of eigenstates of $\{\mathbf{J}^{2},J_{3}\}$  the
generators of rotations are $\mathbf{J}=\bm{\sigma}/2$ and the generators of the
HLG are
\begin{equation}
M_{R}^{ij}=\varepsilon_{ijk}J_{k}=\frac{1}{2}\varepsilon_{ijk}\sigma_{k}%
=\frac{1}{4i}[\sigma_{i},\sigma_{j}],\qquad M_{R}^{0i}=K_{Ri}=iJ_{i}=\frac
{i}{2}\sigma_{i}. \label{MijR}%
\end{equation}
Similarly the generators for the $(0,\frac{1}{2})$ representation are%
\begin{equation}
M_{L}^{ij}=\varepsilon_{ijk}J_{k}=\frac{1}{2}\varepsilon_{ijk}\sigma_{k}%
=\frac{1}{4i}[\sigma_{i},\sigma_{j}],\qquad M_{L}^{0i}=K_{Li}=-iJ_{i}%
=-\frac{i}{2}\sigma_{i}. \label{MijL}%
\end{equation}
According to Eqs.(\ref{BRBL},\ref{rap}) the boost operators read
\begin{align}
B_{R}(\mathbf{p)}  &  =\exp(\bm{\sigma}\cdot\mathbf{n}\frac{\varphi}%
{2})=\frac{E+m+\bm{\sigma}\cdot\mathbf{p}}{\sqrt{2m(E+m)}},\label{BRLJ3}\\
B_{L}(\mathbf{p)}  &  =\exp(-\bm{\sigma}\cdot\mathbf{n}\frac{\varphi}%
{2})=\frac{E+m-\bm{\sigma}\cdot\mathbf{p}}{\sqrt{2m(E+m)}}.\nonumber
\end{align}
These relations allows us to explicitly construct the Weyl states of momentum
$p^{\mu}=(E,\mathbf{p})$, just boosting the rest frame state which have a
canonical form in the $\{|\frac{1}{2},m\rangle\}$ basis. The Weyl spinors with
well defined $J_{3}$ eigenvalues in the rest frame are\bigskip%
\begin{align}
\Phi_{R}(\mathbf{p,+)}  &  \mathbf{=}N(E)\left(
\begin{array}
[c]{c}%
E+m+p_{z}\\
p_{x}+ip_{y}%
\end{array}
\right)  ,\quad\Phi_{R}(\mathbf{p,-)}\mathbf{=}N(E)\left(
\begin{array}
[c]{c}%
p_{x}-ip_{y}\\
E+m-p_{z}%
\end{array}
\right) \label{StatesJ3}\\
\Phi_{L}(\mathbf{p,+)}  &  \mathbf{=}N(E)\left(
\begin{array}
[c]{c}%
E+m-p_{z}\\
-(p_{x}+ip_{y})
\end{array}
\right)  ,\quad\Phi_{L}(\mathbf{p,-)}\mathbf{=}N(E)\left(
\begin{array}
[c]{c}%
-(p_{x}-ip_{y})\\
E+m+p_{z}%
\end{array}
\right)  ,\nonumber
\end{align}
with $N(E)=\left[  2m(E+m)\right]  ^{-\frac{1}{2}}$. Here we used the second
label $\lambda=\pm$ to denote the eigenvalue $\pm1/2$ of $J_{3}$ in the rest frame.

The description of the interactions of spin $\frac{1}{2}$ particles according
to the gauge principle requires to construct first a lagrangian for the free
particle. This is a scalar function and it can be shown that it is not
possible to construct a Lagrangian using only two-dimensional left or right
spinors. This can be exhibited with the mass term for right spinors. It cannot
be of the simplest form
\begin{equation}
\mathcal{L}_{m}\sim\Phi_{R}^{\dagger}\Phi_{R}%
\end{equation}
since it is not Lorentz invariant due to the anti-hermiticity of the boost
operators ($\mathbf{K}=i\mathbf{J}$ with $\mathbf{J}$ hermitian operators)
which yield $\left(  B_{R}(\mathbf{p)}\right)  ^{\dagger}=B_{R}(\mathbf{p)}$.
There is however the possibility to construct the invariant product as
$\overline{\Phi_{R}}\Phi_{R}$ with $\overline{\Phi_{R}}=\Phi_{R}^{\dagger}A$
where $A$ is an operator in the $(\frac{1}{2},0)$ representation space. An
invariant product would require $A^{-1}\left(  B_{R}(\mathbf{p)}\right)
^{\dagger}A=B_{R}(-\mathbf{p})$ which can be rewritten as
\begin{equation}
A^{-1}(-i\mathbf{K})^\dagger A=i\mathbf{K,} \label{aja}%
\end{equation}
but there is no \textit{linear} operator $A$ doing this job for $\mathbf{K}%
=i\bm{\sigma}/2$ . Stated differently, since $B_{R}^{-1}(\mathbf{p}%
)=B_{L}(-\mathbf{p})$, an invariant product requires the adjoint spinor to
transform in the ``left" representation but there is no linear operator
transforming ``left" into ``right" representations since otherwise they would be
equivalent. There is however an\textit{ antilinear} operator doing this job
$A=\xi\sigma_{2}\mathcal{K}$ with $\xi$ a phase and $\mathcal{K}$ the complex
conjugation operator. In any case we must realize that the construction of an
invariant product will always require the combination of ``right" with ``left" states.

The simplest way to accomplish Eq. (\ref{aja}) with a linear operator is to
enlarge the representation space to $(\frac{1}{2},0)\oplus(0,\frac{1}{2})$.
The boost and rotation operators for this reducible representation reads%
\begin{equation}
B(\mathbf{p})=\left(
\begin{array}
[c]{cc}%
\mathbf{\exp(}-i\mathbf{K}_{R}\cdot\mathbf{n}\varphi) & 0\\
0 & \mathbf{\exp(}-i\mathbf{K}_{L}\cdot\mathbf{n}\varphi)
\end{array}
\right)  ,\qquad R(\mathbf{\theta})=\left(
\begin{array}
[c]{cc}%
\mathbf{\exp(-}i\mathbf{J}_{R}\cdot\mathbf{n}\theta) & 0\\
0 & \mathbf{\exp(-}i\mathbf{J}_{L}\cdot\mathbf{n}\theta)
\end{array}
\right)  ,\label{BR}%
\end{equation}
with $\mathbf{J}_{L}=\mathbf{J}_{R}=i\mathbf{K}_{L}=-i\mathbf{K}%
_{R}=\bm{\sigma}/2$ and the generators for $(\frac{1}{2},0)\oplus
(0,\frac{1}{2})$ read
\begin{equation}
M^{ij}=\varepsilon_{ijk}J_{k}=\frac{i}{4}[\gamma^{i},\gamma^{j}]\equiv\frac
{1}{2}\sigma^{ij},\qquad M^{0i}=K_{i}=\frac{i}{4}[\gamma^{0},\gamma^{i}%
]\equiv\frac{1}{2}\sigma^{0i},
\end{equation}
where%
\begin{equation}
J^{i}=\frac{1}{2}\left(
\begin{array}
[c]{cc}%
\sigma_{i} & 0\\
0 & \sigma_{i}%
\end{array}
\right)  ,\qquad K^{i}=\frac{i}{2}\left(
\begin{array}
[c]{cc}%
\sigma_{i} & 0\\
0 & -\sigma_{i}%
\end{array}
\right)  ,\qquad\gamma^{i}=\left(
\begin{array}
[c]{cc}%
0 & -\sigma_{i}\\
\sigma_{i} & 0
\end{array}
\right)  ,\qquad\gamma^{0}=\left(
\begin{array}
[c]{cc}%
0 & 1\\
1 & 0
\end{array}
\right)  .
\end{equation}
Notice that now the boost generators can be written as $\mathbf{K}%
=i\chi\mathbf{J}$ where $\chi$ is the hermitian matrix%
\begin{equation}
\chi=\left(
\begin{array}
[c]{cc}%
1 & 0\\
0 & -1
\end{array}
\right)  .
\end{equation}
Notice also that the eigenstates of this operator are the chiral states
embedded in this larger representation space, i.e., if we define
\begin{equation}
\psi_{R}(p,\lambda)=\left(
\begin{array}
[c]{c}%
\Phi_{R}(\mathbf{p},\lambda\mathbf{)}\\
0
\end{array}
\right)  ,\qquad\psi_{L}(p,\lambda)=\left(
\begin{array}
[c]{c}%
0\\
\Phi_{L}(\mathbf{p},\lambda\mathbf{)}%
\end{array}
\right)  ,\label{psiRL}%
\end{equation}
these are eigenstates of $\chi$ with eigenvalues $+1$ and $-1$ respectively.
We will call this operator as \textit{chirality operator} in the following and 
sticking to the conventional notation we will write it as $\chi = \gamma^{5}$.

The construction of the Lorentz invariant product in the $(\frac{1}%
{2},0)\oplus(0,\frac{1}{2})$ representation space requires the existence of an
operator $A$ fulfilling Eq. (\ref{aja}) . Since for $(\frac{1}{2}%
,0)\oplus(0,\frac{1}{2})$ the boost operator can be written as $-iK^{i}%
\mathbf{=}\frac{1}{2}\gamma^{0}\gamma^{i}$ , it is easy to show that, in order
to satisfy Eq. (\ref{aja}) with a linear $A$, the $\gamma$ matrices must
transform as%
\begin{equation}
A^{-1}\gamma^{0}A=\pm\gamma^{0},\qquad A^{-1}\gamma^{i}A=\mp\gamma
^{i},\label{AA}%
\end{equation}
which has a solution only for the upper sign and is given by $A=\rho\gamma^{0}$
with $\rho$ a phase. Interestingly, from Eqs. (\ref{AB}) it can be shown that
under the discrete symmetry of parity, the representation $(a,b)$ is mapped
onto $(b,a)$ and vice versa, thus the irreps $(\frac{1}{2},0)$ and $(0,\frac
{1}{2})$ are mapped onto each other, meaning that the representation of
parity operator is $\Pi=\zeta\gamma^{0}=A$, a relation valid up to a phase which is 
irrelevant to our construction and will be skipped in the following.
This is consistent with Eq.(\ref{aja}) since also in this case $\left(
B(\mathbf{p)}\right)  ^{\dagger}=B(\mathbf{p})$ and parity is expected to
transform the boost operator as
\begin{equation}
\Pi^{-1}B(\mathbf{p)}\Pi=B(-\mathbf{p}).
\end{equation}
It is necessary to remark that the matrix representation of parity is frame
dependent since it does not commute with the boost operator. If we use the 
boost operator in Eq.(\ref{BR}) with the operators for the chiral boosts in
Eq.(\ref{BRLJ3}), then the representation $\Pi=\gamma^{0}$ is valid only
in the rest frame. The boost operator in this case reads%
\begin{equation}
B(\mathbf{p)}=\left(
\begin{array}
[c]{cc}%
\frac{E+m+\mathbf{\sigma}\cdot\mathbf{p}}{\sqrt{2m(E+m)}} & 0\\
0 & \frac{E+m-\mathbf{\sigma}\cdot\mathbf{p}}{\sqrt{2m(E+m)}}%
\end{array}
\right)  \label{BD}%
\end{equation}
and it can be shown that the parity operator in an arbitrary frame is
\begin{equation}
\Pi(\mathbf{p})\equiv B(\mathbf{p)\Pi}B^{-1}(\mathbf{p)=}\frac{\gamma^{\mu
}p_{\mu}}{m}.
\end{equation}
This reveals the Dirac equation as the eigenvalue equation for parity in the
frame where the particle has momentum $\mathbf{p}$. In the following we will
take the adjoint spinor as $\overline{\psi}\equiv\psi^{\dagger}\Pi=\psi^{\dagger}\gamma^{0}$.

States of momentum $\mathbf{p}$ in the $(\frac{1}{2},0)\oplus(0,\frac{1}{2})$ representation can be
explicitly constructed from states in the rest frame using the boost operator
in Eq. (\ref{BD}). We are interested in states with well defined $J_{3}$. The
most general rest frame spinor satisfying this constriction is
\begin{equation}
w(0,\lambda)=a\psi_{R}(0,\lambda)+b\psi_{L}(0,\lambda),
\end{equation}
with $a^{2}+b^{2}=1$. There are four independent states whose specific form
depend on the values of $a$ and $b$ and many choices for the basis are
possible. The most used basis are Dirac and Weyl ones. The former are parity
eigenstates and correspond to the choice $a=b=1/\sqrt{2}$ for positive parity
and \ $a=-b=1/\sqrt{2}$ for negative parity. The explicit form of these
spinors will be needed below thus we give the explicit result of boosting the 
rest frame states and use the conventional notation $u(\mathbf{p},\lambda)$ and 
$v(\mathbf{p},\lambda)$ for the positive and negative parity states respectively. 
Using the boost operator in Eq. (\ref{BD}) we get
\begin{align}
\label{uv}u\left(  \mathbf{p},+\right)   &  = N\left(
\begin{array}
[c]{c}%
E+m+p_{z}\\
p_{x}+ip_{y}\\
E+m-p_{z}\\
-p_{x}-ip_{y}%
\end{array}
\right)  ,\quad u\left(  \mathbf{p},-\right)  =N\left(
\begin{array}
[c]{c}%
p_{x}-ip_{y}\\
E+m-p_{z}\\
-p_{x}+ip_{y}\\
E+m+p_{z}%
\end{array}
\right)  ,\\
v\left(  \mathbf{p},+\right)   & =N\left(
\begin{array}
[c]{c}%
E+m+p_{z}\\
p_{x}+ip_{y}\\
-E-m+p_{z}\\
p_{x}+ip_{y}%
\end{array}
\right)  ,\quad v\left(  \mathbf{p},-\right)  =N\left(
\begin{array}
[c]{c}%
p_{x}-ip_{y}\\
E+m-p_{z}\\
p_{x}-ip_{y}\\
-E-m-p_{z}%
\end{array}
\right) ,\nonumber
\end{align}
with $N=\left[  4m(E+m)\right]  ^{-1/2}$. These spinors satisfy $\overline
{u}\left(  \mathbf{p},\lambda\right)  u\left(  \mathbf{p},\lambda\right)  =1$
and $\overline{v}\left(  \mathbf{p},\lambda\right)  v\left(  \mathbf{p}%
,\lambda\right)  =-1$.

\section{Second order formalism for spin $\frac{1}{2}$ fermions}

The Pauli Lubanski (PL) operator for the $(\frac{1}{2},0)\oplus(0,\frac{1}%
{2})$ representation is%
\begin{equation}
W_{\alpha}=\frac{1}{2}\varepsilon_{\alpha\beta\rho\mu}M^{\beta\rho}
P^{\mu}=\frac{i}{2}\gamma^{5}\sigma_{\alpha\mu}P^{\mu}%
\label{PLvector}
\end{equation}
and the squared PL operator reads%
\begin{equation}
W^{2}=-\frac{3}{4}\mathcal{T}_{\mu\nu}P^{\mu}P^{\nu}, \label{PLnude}%
\end{equation}
with%
\begin{equation}
\mathcal{T}_{\mu\nu}=g_{\mu\nu}-\frac{i}{2}\sigma_{\mu\nu}.
\end{equation}
The projection over Poincar\'{e} eigensubspaces in the case of a
representation of the HLG containing only one Poincar\'{e} sector reads
\begin{equation}
\frac{P^{2}}{m^{2}}\left(  \frac{W^{2}}{-s(s+1)P^{2}}\right)  \psi=\psi
\end{equation}
and using Eq.(\ref{PLvector}) can be written as
\begin{equation}
\left(  \mathcal{T}_{\mu\nu}P^{\mu}P^{\nu}+m^{2}\right)  \psi=0.
\end{equation}
The antisymmetric part of this tensor does not contribute in the free case but
it becomes active upon gauging. This part is not uniquely fixed by the
Poincar\'{e} projector and the most general form of this tensor consistent with
Poincar\'{e} projection reads
\begin{equation}
T_{\mu\nu}=g_{\mu\nu}-\frac{i}{2}(g +i \xi\gamma^{5}) \sigma_{\mu\nu} \label{tmunu}%
\end{equation}
where $g,\xi$ are free parameters. The corresponding equation of motion is%
\begin{equation}
\left(  T_{\mu\nu}P^{\mu}P^{\nu}+m^{2}\right)  \psi=0, \label{eom}%
\end{equation}
and can be derived from the following lagrangian%
\begin{equation}
\mathcal{L}=\overline{\psi}\left(  -T_{\mu\nu}\partial^{\mu}\partial^{\nu
}+m^{2}\right)  \psi,
\end{equation}
which, using the property $\gamma^0(\Gamma^{\mu\nu})^{\dagger}\gamma^0=\Gamma_{\nu\mu}$ and modulo surface terms can be written in the explicitly Hermitian form%
\begin{equation}
\mathcal{L}=\left(  \partial^{\mu}\overline{\psi}\right)  T_{\mu\nu}%
\partial^{\nu}\psi+m^{2}\overline{\psi}\psi. \label{lag}%
\end{equation}
We remark that unlike the Dirac theory which projects onto well defined parity
subspaces of $(\frac{1}{2},0)\oplus(0,\frac{1}{2})$ , the Poincar\'{e}
projection just select the appropriate spin and mass and it can be applied to
any particle living in the $(\frac{1}{2},0)\oplus(0,\frac{1}{2})$
representation space. In order to gain insight into the structure of this
theory in the following we study the electromagnetic interactions of particles
in this formalism. 

Gauging the lagrangian in Eq. (\ref{lag}) we obtain
\begin{equation}
\mathcal{L}_{int}=-ie\left[  \overline{\psi}T_{\mu\nu}\partial^{\nu}%
\psi-\partial^{\nu}\overline{\psi}T_{\nu\mu}\psi\right]  A^{\mu}%
+e^{2}\overline{\psi}\left(  T_{\mu\nu}\right)  \psi A^{\mu}A^{\nu}.
\end{equation}
which describes the electromagnetic interactions of a particle with charge $-e$
with an external electromagnetic field and we would like to remark that the 
$\xi$ terms in this Lagrangian (see Eq. (\ref{tmunu})), are odd under parity. 

\begin{figure}
\begin{minipage}{0.33\textwidth}
\includegraphics[scale=0.5]{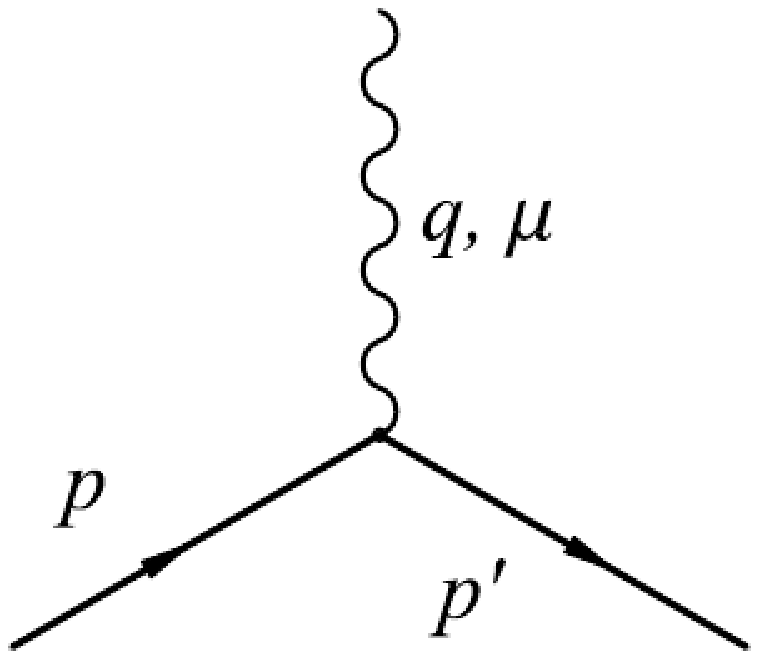}
\end{minipage}
\begin{minipage}{0.33\textwidth}
\includegraphics[scale=0.5]{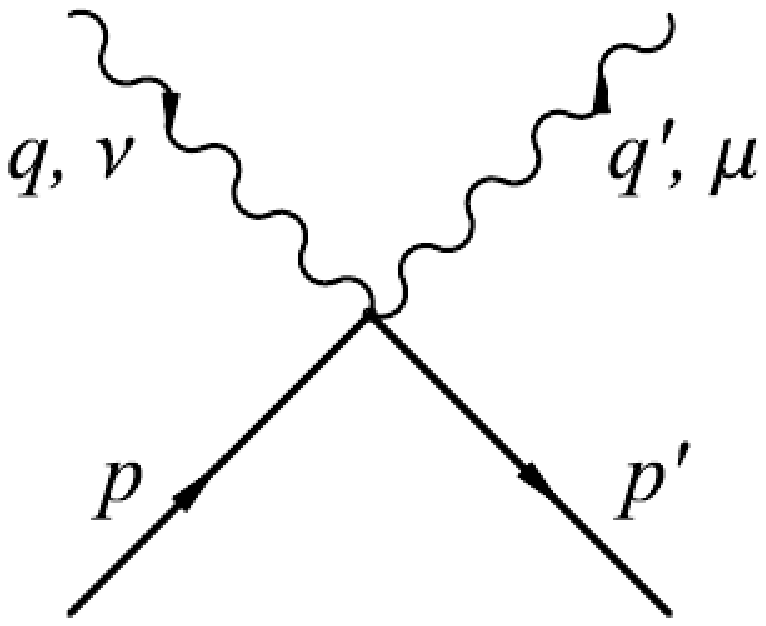}
\end{minipage}
\begin{minipage}{0.32\textwidth}
\includegraphics[scale=0.5]{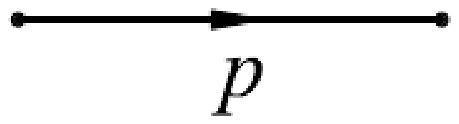}
\end{minipage}\\
\begin{subequations}
\begin{minipage}{0.33\textwidth}
\begin{equation*}
ie\left(  T_{\mu\nu}p^{\nu}+T_{\nu\mu}p^{\prime\nu}\right)  \equiv ieV_{\mu}(p^{\prime},p),
\end{equation*}
\end{minipage}
\begin{minipage}{0.33\textwidth}
\begin{equation*}
-\frac{ie^{2}}{2}(T_{\mu\nu}+T_{\nu\mu}),
\end{equation*}
\end{minipage}
\begin{minipage}{0.32\textwidth}
\begin{equation*}
i\mathcal{P}(p)=\frac{i}{p^{2}-m^{2}},
\end{equation*}
\end{minipage}
\end{subequations}
\caption{\label{FR} Feynman rules for the second order formalism}
\end{figure}

The Feynman rules arising from this lagrangian are depicted in figure \ref{FR}.
The vertex function satisfy the Ward-Takahashi identity independently of the
value of the parameters $g, \xi$%
\begin{equation}
\left(  p^{\prime}-p\right)  ^{\mu}V_{\mu}(p^{\prime},p)=T_{\nu\mu}%
p^{\prime\nu}p^{\prime\mu}-T_{\mu\nu}p^{\mu}p^{\nu}=\mathcal{P}^{-1}%
(p^{\prime})-\mathcal{P}^{-1}(p). \label{WT}%
\end{equation}
In general, Lorentz covariance allows us to write the one particle state as
$\psi(x)=w(p,\lambda)e^{-ip\cdot x}$ where $w(p,\lambda)$ is a four-component
spinor. The explicit form of these spinors depends on the nature of the
particle we want to describe and can be constructed from first principles
solving the Lorentz algebra as we have done above. In order to compare with the 
results of the conventional Dirac theory, in the next section we calculate
Compton scattering off spin 1/2 fermions with well defined parity in this formalism.

\section{Compton scattering}

\begin{figure}
\includegraphics[scale=0.5]{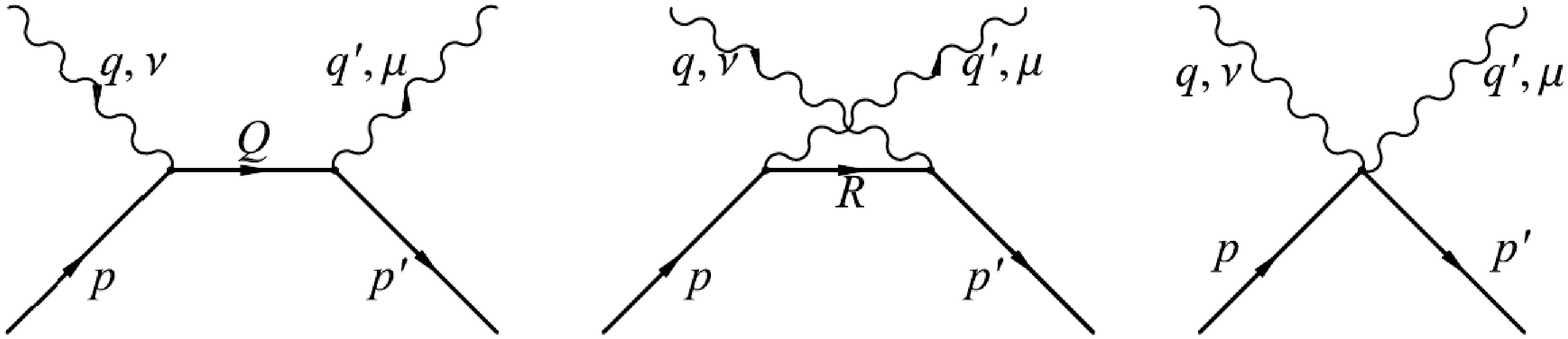}
\caption{\label{CS} Feynman diagrams for Compton scattering}
\end{figure}

In this section we calculate Compton scattering of a spin $1/2$ particle in
the second order formalism. The Feynman rules for Compton scattering are given
in Fig. \ref{CS}.
A straightforward calculation yields the following invariant amplitude%
\begin{equation}
-i\mathcal{M}=e^{2}\overline{w}(p^{\prime})\Gamma_{\mu\nu}w(p)\varepsilon
^{\mu}(q^{\prime})\varepsilon^{\nu}(q)
\end{equation}
with%
\begin{equation}
\Gamma_{\mu\nu}=\frac{V_{\mu}(p^{\prime},Q)V_{\nu}(Q,p)}{Q^{2}-m^{2}}%
+\frac{V_{\nu}(p^{\prime},R)V_{\mu}(R,p)}{R^{2}-m^{2}}-\Gamma_{\mu\nu}%
-\Gamma_{\nu\mu}%
\end{equation}
and $Q=p+q=p^{\prime}+q^{\prime}$, $R=p-q^{\prime}=p^{\prime}-q$. \ Using the
Ward-Takahashi identities it easy to show that this amplitude is gauge
invariant. Squaring this amplitude we obtain%
\begin{equation}
|\overline{\mathcal{M}}|^{2}=\frac{e^{2}}{4}tr\left[  \Gamma_{\mu\nu
}\mathcal{S}(p)\overline{\Gamma}^{\mu\nu}\mathcal{S}(p^{\prime})\right]
\label{amp2}%
\end{equation}
where $\overline{\Gamma}^{\mu\nu}=\Pi\left(  \Gamma^{\mu\nu}\right)
^{\dagger}\Pi$ and $\mathcal{S}(p)$ denotes the sum over polarizations of the
spin $1/2$ particle%
\begin{equation}
\mathcal{S}(p)=\sum_{\lambda}w(p,\lambda)\overline{w}(p,\lambda)\text{.}%
\end{equation}
Here is where the specific nature of the particle beyond its mass and spin
enter the formalism. We will calculate Compton scattering off particles with
well defined parity in order to compare our results with those of the Dirac
formalism. The corresponding states were calculated above by solving the 
HLG algebra and are given in Eq. (\ref{uv}). Either boosting the rest frame 
parity projector or using the explicit form of the states it is easy to show that%
\begin{equation}
\mathcal{S}(p)=\sum_{\lambda}u(p,\lambda)\overline{u}(p,\lambda)=
\frac{\not p+m}{2m}.
\end{equation}
Inserting this projector in Eq. (\ref{amp2}) we obtain%
\begin{equation}
|\overline{\mathcal{M}}^{2}|  = \frac{e^{4}}
{16m^{2}\left( m^{2}-s\right)^{2}\left(  m^{2}-u\right)^{2}}\sum_{n=0}^{5} m^{2n}h_{2n}(s,u)
\label{M2}
\end{equation}
where
\begin{eqnarray}
h_{0}&=& 2 [(g-2)^2+\xi^{2}]\left(g^2+\xi^2+2\right) (s+u) s^2 u^2, \nonumber \\
h_{2}&=&-s u [g^4 (3 s+u) (s+3 u)-8 g^3 \left(s^2+4 s u+u^2\right)+2 g^2
   \left( 2 \left(s^2+8 s u+u^2\right)+\xi ^2 (3 s+u) (s+3 u)\right) \nonumber \\
   &&+8 g\left(-\xi^2 \left(s^2+4 s u+u^2 \right)
   +2 \left(s^2+u^2\right)\right)+12 \xi^2 \left(s^2+4 s u+u^2\right)+
   \xi^4 (3 s+u) (s+3 u)-32 s u ] ,\nonumber \\
h_{4}&=&(s+u) \left[ \left(\left(g^2+\xi ^2\right)^2+16\right)(s^2+u^2)
  +2 s u \left( 2 (g (5 g-4)+6) \xi^2+g ((g-2) g (5 g+2)+48)+5 \xi^4\right)\right],
   \nonumber \\
h_{6}&=& -\left( s^2+u^2\right) \left[ 2 g (3 g+4) \xi^2+g (g+2)^2 (3 g-4)+3 \xi^4+4
   \left(-3 \xi^2+8\right)\right] \nonumber \\
   &&-2 s u \left[ 2 g (5 g+4) \xi^2+g (g (g (5 g+8)-20)+48)+5 \xi^4 - 12 
   \xi^2+64\right],\nonumber \\
h_{8}&=&2 \left[g^4+12 g^3-2 g^2 + 2 (g (g+6)-9)\xi^2-40 g+\xi^4-24\right](s+u)\nonumber \\
h_{10}&=&-16 g^3-8 g^2-16 g \xi^2+96 g+24 \xi^2+160,
\label{hs}
\end{eqnarray}
and $s,u$ are the Mandelstam variables.
We would like to remark that the squared amplitude is symmetric under the exchange $s\leftrightarrow u$ and in the case $g=2,~\xi=0$, Eq. (\ref{M2}) reduces to the
conventional Dirac result
\begin{equation}
|\overline{\mathcal{M}}^{2}|_{D}   = \frac{4 e^{4}}{2\left(  m^{2}-s\right)
^{2} \left(  m^{2}-u\right)  ^{2}} \left[ 6 m^{8}-\left(  3 s^{2}+14 u s+3
u^{2}\right)  m^{4}
  +\left(  s^{3}+7 u s^{2}+7 u^{2} s+u^{3}\right)  m^{2}-s u \left(
s^{2}+u^{2}\right)  \right].
\label{Dirac}
\end{equation}

In the laboratory system the Mandelstam variables are given by
\begin{equation}
s=(p+q)^{2}=m(m+2\omega),\qquad
t=(q^{\prime}-q)^{2}=-2\omega\omega^{\prime}(1-\cos\theta),\qquad
u=(p-q^{\prime})^{2}=m(m-2\omega^{\prime}),
\label{manlab}
\end{equation}
where $\omega$, $\omega^{\prime}$ denote the incoming and outgoing photon
energy respectively which are related as
\begin{equation}
\omega^{\prime}=\frac{m\omega}{m+\omega(1-\cos\theta)}, \label{wp}%
\end{equation}
with $\theta$ denoting the angle of the outgoing photon with respect to the
direction of the incident one.

The angular distribution of the photons in this frame is
\begin{equation}
\frac{\textup{d}\sigma}{\textup{d}\Omega}=\frac{1}{4(4\pi)^{2}}\frac
{|\overline{\mathcal{M}}|^{2}}{m^{2}}\left(  \frac{\omega^{\prime}}{\omega
}\right)  ^{2}. \label{comptondcs}%
\end{equation}
Using Eqs. (\ref{M2},\ref{manlab},\ref{wp},\ref{comptondcs}) we obtain
\begin{eqnarray}
\frac{d\sigma(g,\xi,\eta,x)}{d\Omega}  &=&\frac{r_c^2}{64 ((x-1) \eta -1)^3} 
[2 x \eta  \left(\eta  \left(g^4+2 g^2 \left(\xi ^2-1\right)+8 g+\xi ^4+8\right)+16\right)
\nonumber \\
&+&x^2 \left(\eta
   ^2 \left(g^4-8 g^3+2 g^2 \left(\xi^2+4\right)-8 g \xi^2+\xi ^4+12 \xi^2-16\right)-32 \eta -32\right) \nonumber \\
   &-&\eta^2 \left(3
   g^4-8 g^3+g^2 \left(8+6 \xi^2\right)-8 g \xi^2+3 \xi^4+12 \xi^2+16\right)+4 x^3 \eta  \left((g-2)^2 \eta +8\right)-32 \eta -32],
\label{angdist}
\end{eqnarray}
where $\eta=\omega/m$ and $x=\cos\theta$ and $r_{c}=\alpha/m$ denotes the
classical radius of the spin 1/2 particle.

In the classical limit ($\eta<< 1$) the differential cross section is
independent of the value of $g,\xi$ 
\begin{equation}
\frac{d\sigma(g,\xi,\eta,x)}{d\Omega}|_{\eta<<1}=
\frac{r_{c}^{2}}%
{2}\left(  x^{2}+1\right)  +r_{c}^{2}\left(  x^{3}-x^{2}+x-1\right)  \eta,
\end{equation}
and to leading order we obtain the Thomson angular distribution. The next to leading 
order term is also independent of the parameters $g,~\xi$ a fact also encountered in 
the treatment of spin $1$ and $3/2$ in the projector formalism. It can be shown that these are also the leading terms in scalar electrodynamics. The dependence on the parameters $g,~\xi$ in Eq.(\ref{angdist}) appears at order $\eta^2$. 

In the forward direction ($x=1$) the differential cross section is independent of
the photon energy and of the value of $g,~\xi$ taking the value
\begin{equation}
\frac{d\sigma(g,\xi,\eta,x)}{d\Omega}_{forward}=r_{c}^{2}.
\end{equation}
Finally, for $x\neq 1$, at high energies it vanishes independently of the value of $g,\xi$
\begin{equation}
\frac{d\sigma(g,\xi,\eta,x)}{d\Omega}|_{\eta\rightarrow\infty}=0.
\end{equation}
Integrating the angular distribution we obtain the total cross section as%
\begin{eqnarray}
\sigma(g,\xi,\eta) &=&\frac{3\sigma_{T}}{256\eta^{3}(2\eta+1)^{2}}\left\{
2 \eta  \left(4 \left(3 g^2-12 g+76\right) \eta  \right.\right. \nonumber \\
&+&  4 \eta^4 \left(g^4-4 g^3+2 g^2 \left(\xi^2+3\right)-4 g \left(\xi^2+2\right)
     +\xi^4+6 \xi^2+8\right)  \nonumber \\
&+& \eta^3 \left(3 g^4-24 g^3+g^2 \left(76+6 \xi^2\right)-8 g \left(3 \xi^2+26\right)
 +3 \left(\xi^4+12 \xi^2+96\right)\right)\nonumber \\
&+& \left. \eta ^2 \left(g^4-8 g^3+2 g^2 \left(\xi^2+28\right)-8 g
   \left(\xi^2+24\right)+\xi^4+12 \xi^2+496\right)+64\right) \nonumber \\
&-&  (2 \eta +1)^2 \log (2 \eta +1) \left[4 \left(3 g^2-12
   g+28\right) \eta \right. \nonumber \\
&+& \left.\left.\eta ^2 \left(g^4-8 g^3+2 g^2 \left(\xi^2+10\right)
   -8 g \left(\xi^2+6\right)+\xi^4+12 \xi^2+32\right)+64\right] \right\} ,
\end{eqnarray}
where $\sigma{_{T}}$ denotes the Thomson cross section
\begin{equation}
\sigma_{T}=\frac{8}{3}\pi r_{c}^{2}.
\end{equation}
In the classical limit we recover the Thomson value
\begin{equation}
\sigma(g,\xi,\eta)|_{\eta\rightarrow0}=\sigma_{T}%
\end{equation}
independently of the value of $g,~\xi$ . At high energies we get
\begin{equation}
\sigma(g,\eta)|_{\eta\rightarrow\infty}=\frac{3}{128}\left[(g-2)^{2}+\xi^2\right]\left(
g^{2}+\xi^2+2\right)  \sigma_{T}.
\end{equation}
We remark that the total cross section vanishes at high energies only for
$g=2,~\xi=0$, otherwise it takes a finite value.

Summarizing the so far obtained results, we developed a second order
formalism for the description of fermions based on Poincar\'e projectors 
which contains two  free parameters $g,~\xi$, the later yielding odd parity terms
in the lagrangian. It is important to remark that
unlike the first order Dirac theory, the second order formalism fixes only the
mass and spin of the particle but not any other property. When applied to
states with well defined parity, it still leaves us the freedom to choose the
value of the gyromagnetic factor $g$ although for the specific value $g=2$ it
yields the same results than Dirac theory. 

The forward differential cross section takes the fixed value $r_{c}^{2}$ independently of the photon energy and of the values of the parameters. Interestingly, the same result is obtained for a scalar particle in the conventional scalar electrodynamics and for a 
spin 1 particle in the Standard Model. A similar result is obtained 
in the second order formalism for the vector\cite{Napsuciale:2007ry} and spinor-vector 
representation \cite{DelgadoAcosta:2009ic} but only if $g=2$ and the odd parity 
structures vanish. We must mention that this result is not obtained for spin 3/2 
particles in the Rarita-Schwinger formalism.  On the
light of this result, it would be interesting to revisit the sum rules results
that yields $g=2$ at tree level as a fundamental quantity for spin 1/2 and
which are based on the vanishing of the forward amplitudes for Compton
scattering at high energies \cite{Weinberglectures}.

Another interesting possibility is the application of the formalism to the
description of the electromagnetic interactions of baryons at low energies.
From the effective field theory point of view, the magnetic moment of a baryon
is a low energy constant which encodes  the information of the structure of
the baryon which is irrelevant at low energies. Clearly, the value $g=2$
predicted by the Dirac theory is not the most appropriate starting point for
the corresponding description. For instance, in the case of the proton the 
gyromagnetic factor is $g_{P}=2+\kappa=5.58$ and the "anomalous" magnetic moment
of a proton with polarization $\lambda =1/2$, $e\kappa/4m_{P}$ is even larger than the "natural" magnetic moment $e/2m_{P}$.

As an example of the possible application of the second order formalism to the 
effective field theory description of hadron properties below we
compare the results of this formalism for the differential cross
section of the proton Compton scattering at low energies with experimental
results recently obtained at MAMI using the TAPS detector system setup \cite{OlmosdeLeon:2001zn}.
\begin{figure}
\includegraphics[scale=0.8]{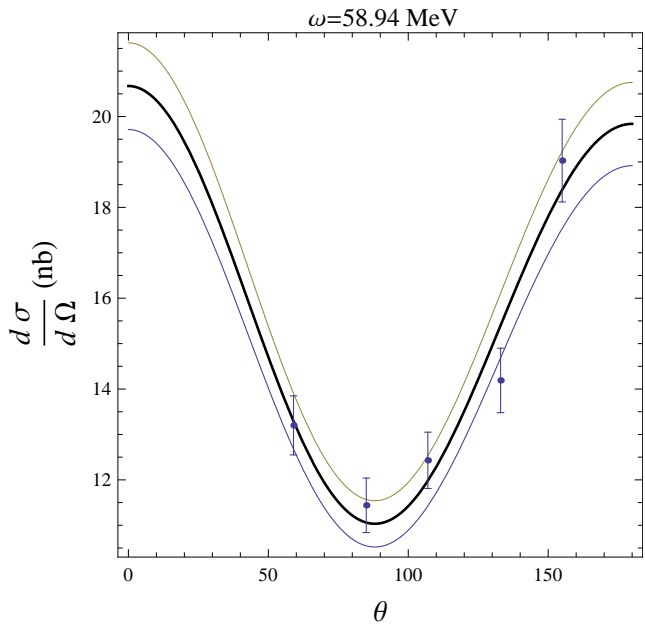}
\caption{\label{angdistplot} Angular distribution of the outgoing photons for 
$\omega =58.94~ MeV$. The central curve is obtained from the fit to Eq. \ref{angdist} with $\xi=0$. The upper and lower curves correspond to the one sigma region of the fit.}
\end{figure}
 We use the data for the angular distribution with an incident photon energy $\omega =58.94~ MeV$. We fix $\xi=0$  and fit the parameters $r_{c}$, and $g$ to the data in Table 2 of \cite{OlmosdeLeon:2001zn}. We obtain the central values $r_{c}=14.37 \times 10^{-18} m$, $g_{P}=5.580$ and a $\chi^2/dof=1.09$. The resulting curve is shown in figure \ref{angdistplot} with the one sigma region for the parameters. If we perform a fit fixing $r_{c}$ to its theoretical value $r_{c}=\alpha/m_{P}=15.34\times 10^{-18} m$ we obtain a worse $\chi^2/dof=1.73$, $g_{P}=4.18$ and $\xi=5.04\times 10^{-6}$. In any event, it is clear that the description of the proton electromagnetic properties at low energies requires a more elaborate analysis. In particular it is necessary to include the polarizability of the proton which is strong and due mainly to the $\Delta (1232)$ resonance. This is more transparent when we analyze the experimental data for the differential cross section at a fixed angle as a function of the incident photon energy. For example, taking 
 $r_{c}=14.37 \times 10^{-18} m$ and $g_{P}=5.580$, from Eq. (\ref{angdist}) we obtain the results shown in  figure \ref{ds} together with data form MAMI \cite{OlmosdeLeon:2001zn} 
 for fixed $\theta=107^{\circ}$.
\begin{figure}[ptb]
\includegraphics[scale=0.8]{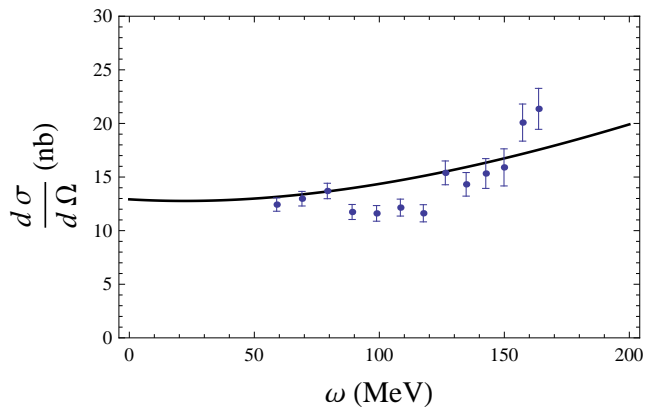}%
\caption{ Differential cross section for Compton scattering off
protons at fixed angle $\theta=107^{\circ}$ as a function of the incoming photon energy
$\omega$. Data is taken from Table 2 of \cite{OlmosdeLeon:2001zn}. The solid line is the
prediction of the second order formalism when we use the central values for $r_{c}$ and $g_{P}$ in the fit to the angular distribution in figure \ref{angdistplot}.}
\label{ds}
\end{figure}
We obtain the correct description at very low energy. For photon energies above 80 MeV the effects of the $\Delta (1232)$, which has a width of the order of 120 MeV, start being visible. We remark that a complete description of the proton electromagnetic properties 
at low energy is beyond the scope of this work and we just aim to show that such
description is possible in the second order formalism and that, from the effective 
theory point of view,  it is a more natural starting point.

Finally we would like to stress that our formalism differs from the Feynman-Gellmann 
second order formalism for fermions in \cite{Feynman:1958ty}. Indeed, Feynman-Gellman description is based on the decomposition of Dirac states into Weyl states and the 
rewriting of the conventional Dirac theory in this basis. As a consequence, this 
formalism is still appropriate to describe particles with well defined parity, although 
in a non conventional basis. The physics under gauging is the same as that of 
Dirac theory and in particular a Feynman-Gellman fermion has also $g=2$. Clearly, our
formalism generalize these results. 

\section{Summary}

In this work we develop a second order formalism for spin 1/2 fermions based
on the projection over Poincar\'{e} invariant subspaces in the $(\frac{1}%
{2},0)\oplus(0,\frac{1}{2})$ representation of the homogeneous Lorentz group.
The theory fixes only the mass and spin of the particle but not any other
property and in general contains two free parameters, $g$ and $\xi$, which 
cannot be fixed from Poincar\'{e} projectors, the latter being related to 
odd-parity Lorentz structures. Gauging the formalism we obtain a gauge invariant
second order description for the electromagnetic interactions of a spin 1/2
fermion. We calculate Compton scattering off a spin 1/2 fermion in this formalism. 
In the particular case $g=2,\xi=0$ and for states in $(\frac{1}{2},0)\oplus(0,\frac{1}{2})$ with well defined parity reduces our results reduce to those of the conventional 
Dirac theory. In the general case we find the correct classical limit and a finite
value $r_{c}^{2}$ for the forward differential cross section, independently of 
the photon energy and of the value of the parameters $g$ and $\xi$. 
Interestingly, the differential cross section vanishes at high energies 
for all $g,~\xi$ except in the forward direction.  For the total cross section at 
low energies we obtain the Thomson value for all $g,~\xi$ and at
high energies it vanishes only for $g=2,~\xi=0$, reaching a finite value otherwise. 
We argue that from the effective theory point of view, the formalism is more 
appropriate than Dirac theory as a starting point for the description of 
electromagnetic properties of baryons and illustrate the point analyzing the 
low energy electromagnetic properties of the proton and comparing with recent 
experimental results.

\begin{acknowledgments}
Work supported by CONACyT-M\'{e}xico under project CONACyT-50471-F. We thank Mariana Kirchbach for useful comments and suggestions. 
\end{acknowledgments}

\end{document}